\documentclass[11pt]{article}
\usepackage{times,aaspp4,astrobib,epsf}

\input{epsf}

\def\lsim{\hbox{ \rlap{\raise 0.425ex\hbox{$<$}}\lower 0.65ex\hbox{$\sim$} }}
\def\gsim{\hbox{ \rlap{\raise 0.425ex\hbox{$>$}}\lower 0.65ex\hbox{$\sim$} }}

\def\etavec{\hbox{$\eta\hskip-6.0pt\eta$}}
\def\zetavec{\hbox{$\zeta\hskip-6.0pt\zeta$}}
\def\Thetavec{\hbox{$\Theta\hskip-8.0pt\Theta$}}

\def\del{\partial}

\def\opeqn{\begin{equation}}
\def\cleqn{\end{equation}}

\begin{document}

\title{ Superluminal Caustic is Just a Common Misconception: \ \ A Comment 
\\ on astro-ph/0001199  by Zheng Zheng and Andrew Gould}
\author{ Sun Hong Rhie }
%\address{ Physics Department, University of Notre Dame, Notre Dame, IN 46556 }
\affil{Physics Department, University of Notre Dame, Notre Dame, IN 46556}

\centerline{(\today)}
\begin{abstract}

When angular objects in lensing are considered as linear objects,
interesting phenomena start happening.  Tachyonic caustics are one example.
We review that the intrinsic variables of the lens equation  are angular
variables. We argue that the ``fast glance effect" of a caustic curve that is
far away from lenses does not share the physical bearing of the well-known 
(apparent) superluminal motion.  There is no dbout that it would be a useful 
exercise to study the null geodesics in the metric of, say, a rapidly 
rotating black hole binary.   Lienard-Wiechert potentials ($A_\mu$) satisfy
Maxwell's equations in Minkowski space.  Authors' claim that  swapping
$eQ$ and $GM$ makes the time component ($A_0$) of the Lienard-Wiechert potentials
into "the gravitational analog" that governs the behavior of the null geodesics
near a relativistic binary system seems to be unfounded.

\end{abstract}

\newpage

It has been a long controversy in the smoky backroom  where non-smoking 
jurors shred papers and throw verdicts, where the caustics are.   The 
controversy hits home often because we are looking for Bruno's planets.
Caustics matter greatly in the field of microlensing planet searches. 
We wrote in a paper on the discovery of evidence of a low mass planet
\cite{98blg35} that "a single lens (stellar lens only) has a point caustic
at the position of the lens."  Or, is the point caustic behind the position
of the lens at the projected position of the lens?  May we draw the caustic
curve and the critical curve in the same plane as we usually do?  
Is it a law or a rule that  
caustics (onto which the critical curve in the image plane  is mapped 
under the lens equation) lie in the source plane?  
Or, is it a matter of definitions and conventions?
What is the image plane?  What is the source plane?
Does any of them (including what we wrote in the paper mentioned)  have any
merit or physical relevance?   What is a superluminal caustic?  We realize
that we just found in the phantom of tachyonic caustics an important clue
to the mysterious misunderstanding behind the controversy. 
Discoveries advance science. So does reasoning. 
Here we investigate the misconception of superluminal caustics 
\cite{whatsupernatural} as an attempt to straighten out the wrinkles 
on caustics.    We only need to borrow a pinch of salt from a way of 
thinking in science which may have been popularized by Einstein: 
\ gedenken experiment.   

\begin{itemize}
\item
  \underline{\Large  Much Ado About Nothing }
\end{itemize}

Andrew Gould writes in a recent article \cite{whatnatural} (G1421 from here on)
that  ``the geometry of point-lens microlensing is so simple that students
can derive all the basic results in a few hours."  The abstract of G1421 
starts with a paragraph,  ``if the standard microlensing geometery is
inverted so that the Einstein ring is projected onto the observer plane
rather the source plane, ... ."    What could be the source plane and 
the observer plane referred to as in G1421?   
In lensing, there are three basic objects, which we may refer to as
``lensing trio" in this article.  They are radiation  an emission source, 
a lensing object, and an observer \cite{schechter}.  
The lensing trio fall more or less on a 
line in the radial direction, namely, the line of sight (of the observer
of the source star in the absence of the intervening lensing object).      
Given a geometric line, one can imagine an infinitely 
many planes that are perpendicular to the line.   
One may refer to the plane that passes through
the radiation emission source as the source plane and the plane that passes
through the observer as the observer plane.    Where is the Einstein ring?
Equation (1) in G1421 indicates that the Einstein ring lies on the plane that 
passes through the lensing object.   
Zheng and Gould (2000; ZG1199 from here on)
refer to the plane through the lensing object  as  the lens plane.  

Physics lies in relations not in  nomenclatures, and it will be most harmonious
if the nomenclatures faithfully represent the relations.  
One of the governing relations in lensing is the so-called lens equation, 
and ZG1199 writes the lens equation in terms of the variables 
defined on the aforementioned source plane and lens plane \cite{sweiss86}.
\opeqn
  {\etavec} = {D_s\over D_1}~{\zetavec} - D_2~\Thetavec({\zetavec}) \ , 
\label{eqZG11}
\cleqn
where $D_1$ and $D_2$ are the distances from the lens to the observer
and to the source, and $D_s = D_1 + D_2$.  ZG1199 describes the transverse 
(or 2-d) position variables ${\etavec}$ and ${\zetavec}$  as follows:
A photon comes from point ${\etavec}$ in the source plane and hits point   
${\zetavec}$ in the lens plane.   At this point, one may wonder if ${\etavec}$
must be  the variable for the source position (since it is said to be in the 
source plane) and ${\zetavec}$ must be the variable for the lens position 
(since it is said to be in the lens plane).  
In order to understand the significance of the variables 
${\etavec}$ and ${\zetavec}$,  we consider a  gendenken experiment:    
we reduce the mass $M$ gradually by taking away one atom at a time.   
A photon from point  ${\etavec}$ on the plane at a distance $D_s$ from the 
observer must hit point ${\zetavec}$ on the lens plane  such that 
lens equation (\ref{eqZG11}) with decreasing mass $M(t)$ is satisfied.   
When the last atom is taken away,  there is no lensing mass, and  there is
no lens plane.   The plane at a distance $D_1$ from the observer is just one
of the infinitely many planes that are perpendicular to the line of sight.
On the other hand, $D_1$ retains its significance in equation (\ref{eqZG11}).
When there is no lensing mass, $\Thetavec$ vanishes, and the lens equation 
reads as follows. 
\opeqn
  \etavec = {D_s\over D_1}~{\zetavec}   
\label{eqNolens}
\cleqn
Since there is no lensing mass, the distance $D_1$ does not have any physical 
relevance and should not show up in  equation (\ref{eqNolens})
with any significance.   But it does.    We only  know that that is 
where we used to have a lensing mass.  Is it some sort of hysteresis?
Then, it could be where we are thinking of putting a  lensing mass because   
we can start piling atoms the very next moment.   
Then, of course, it  could be just a plane we are
thinking of for no reason, perhaps out of boredom.   Or, perhaps, ${\zetavec}$
is not a most representative variable for the governing equation of the 
lensing behavior.  What equation (\ref{eqNolens}) says
is that  the transverse vector ${\etavec}$ on the source plane at a distance  
$D_s$ from the observer and the transverse vector ${\zetavec}$ on an arbitrary
plane at a distance $D_1$ from the observer are parallel and scale with the 
distances from the observer.     What should strike us by now is that
the perspective of the observer, that is all there is in the lens  equation 
with zero lens mass.    So,  we divide the transverse vectors by the 
distances of the planes, and it becomes very clear what the equation
must mean.   Let $\vec\alpha\equiv \zetavec / D_1$ and
$\vec\alpha_s\equiv \etavec / D_s$,   then 
\opeqn
 \vec\alpha = \vec\alpha_s \ .  
\label{eqAlphaZeromass}
\cleqn
A radiation source that would be seen at an angular position $\vec\alpha_s$ 
by an observer in a flat space  is seen by the observer 
at an angular position $\vec\alpha$ that has the same value 
as $\vec\alpha_s$  when there is nothing to change the light ray from that in
the Minkowski space.   Then, it is clear that the pair of variables 
$\{\etavec, \zetavec\}$ simply are not the proper pair of variables for the 
lens equation.  It is not that one can not write the equation in terms of
the variables $\{\etavec, \zetavec\}$, 
but that they oscure the conceptual underpinning of the lens 
equation.   It is also clear that the variable $\vec\alpha$ or 
$\zetavec=D_1 \vec\alpha$ is a variable for the positions of the images
not the lenses.  This fact seems to have generated  another misconception
that images should lie on the lens plane.   This misconception seems to have
produced a  corollary that images can be in two different lens planes (when 
the lensing involves double scattering) as the audience was told by an invited 
speaker in a recent lensing meeting.   

Now, we put back in the lensing mass, say, at $D_1$.
The (2-d) scattering angle $\Thetavec$  is a function of a dimensionful
constant $GM$  as well as of ${\zetavec}$: ~$\Thetavec({\zetavec}; GM)$,
and  it works out to be dimensionless as it should.  
So, we can divide equation (\ref{eqZG11}) by $D_s$ to write it
in terms of the pair of angular variables $\vec\alpha$ and $\vec\alpha_s$.
\opeqn
  \vec\alpha = \vec\alpha_s + {D\over D_1}\Thetavec \ ,
\label{eqLeqAlpha}
\cleqn
where $D$ is the reduced distance
\opeqn
{1\over D} = {1\over D_1} + {1\over D_2}  \ .
\cleqn
A radiation source that would be seen at an angular position $\vec\alpha_s$
by an observer in Minkowski space  
is seen by the observer at an angular position $\vec\alpha$ which is shifted
from $\vec\alpha_s$ by a fraction of the (2-d) scattering angle $\Thetavec$.
The reduced distance is no bigger than the smaller of $D_1$ and $D_2$ as
is familiar from the reduced mass in mechanics. Thus, $D/D_1 \leq 1$, and
the angular shift between $|\vec\alpha-\vec\alpha_s| \leq |\Thetavec|$. 
(The equality holds when $D_1 =0$, which is not exactly a physical situation.
That is because of the hidden assumption of the lens equation that the 
observer is supposed to be asymptotically far away from the lensing mass
as it should be clear in the following section.)   

An emblematic analogue to this interpreational issue of the lens equation
hinging on ``trivial math" and ``deeper physics" 
 would be the case of the cosmological constant which
may have been the biggest blunder of the man of the 20th century (TIME) 
but likely a necessity for the millenials.   
As a desperate effort to stop the universe
from expanding without introducing negative density or pressure, Einstein 
modified his equations in 1917 and introduced a new fundamental constant, 
the so-called cosmological constant.  The new equation read as follows
\cite{einstein,sweinberg}.   
\opeqn
    G_{\mu\nu} - \Lambda g_{\mu\nu} = -8\pi G T_{\mu\nu}  \ ,  
\label{eqLambda}
\cleqn
where $ G_{\mu\nu}$ and $ T_{\mu\nu}$ are Einstein and energy-momentum 
tensors, and $g_{\mu\nu}$ is the metric tensor.   The LHS is the geometry,
the RHS is the matter content,  and the Einstein field equation tells us how
the geometry of space time interacts with the content of the matter in the
space time.   What is curious about the cosmological term is that it does
not vanish when the space time is flat.   It is a blow to the fundamental 
notion of Einstein gravity one might have just convinced oneself to accept: 
\ gravitational interaction is an experience of the curvature of space time. 
The cosmological term is a non-curvature term of the geometry that participates
in  governing the grvaity as it is written in the LHS of the gravitational 
field equation.   We may consider the $\Lambda$-term as a part of the energy 
momentum tensor  and write the gravitational field equation as follows.   
\opeqn
    G_{\mu\nu} 
   =  -8\pi G \left(T_{\mu\nu} - {\Lambda\over 8\pi G}~ g_{\mu\nu}\right)   
\label{eqLambdaVac}
\cleqn
The transition from equation (\ref{eqLambda}) to equation (\ref{eqLambdaVac})
is less than trivial mathematically, but it requires a profound change in the 
frame of physical understanding.  The $\Lambda$-term as an energy momentum
tensor features  negative pressure, which Einstein axiomized to avoid and 
was one of the very reasons why he devised the $\Lambda$-term  in equation
(\ref{eqLambda}) in the first place.   
It took developement of Goldstone bosons, renormalization and brief 
marriage of particle physics and condensed matter physics, Grand Unification 
Theories (GUT),  experience of rich vaccuum structures (with endless 
parade of scalar fields as yet to be discovered or refuted) and phase 
transitions in the early universe,  monopole problem,
horizon problem,  dark matter problem, inflation theories,  structure seed
problem, fine-tuning problems, topological defects,  etc, for the negative
pressure to find its natural position in human intellectural domain.  Now,
no one doubts the physical relevance of the cosmological constant as the
the vaccuum energy density -- or the ``zero point energy" of the (future)
quantum gravity,  even though it is a good question how big it is, or whether
it is a constant.  In astrophysical practice,  it is simply passed as a stuff 
with stiff equation of state: $p = -\rho$ that can overcome the self-gravity 
of ordinary matter and make the universe fly apart.  
In fact,  high-z supernovae searches have found evidence of accelerating 
universe \cite{SNcp,highzSN}.   

So, we emphatically conclude that the intrinsic variables in lens 
equations in (\ref{eqZG11}) and (\ref{eqLeqAlpha}) are angular variables.  
This obvious conclusion is self-evident from the beginning  if we derive 
the lens equation of a point mass $M$ starting from the first principle 
(simply following a standard textbook on general relativity).   An equation
that relates two angular variables, that is the first thing we get.  
The others are all simple derived quantities, and there is no 
misunderstanding of what is what.  If it takes a few hours for
students to do so as Gould (G1421) testifies, it should take less than a
few days for practitioners to derive the lens equation and less than a few 
minutes to read one.  So, we write out the derivation in the following section 
as an effort to abolish the ground for the mysterious understanding that seems 
to perpetuate even more mysterious controversies.  In fact, it is cathartic 
to go through the derviation of the  lens equation once.
The Einstein field equations are non-linear and can not be solved in general,
but there are some exact solutions.  And, the lens equation of a point mass
is derived from one of those exact solutions, and that the simplest one
(despite the problem of no global time killing vector and of nontrivial topology).
It is such a great assurance to be backed by an analytic derivation from 
an exact solution. 

Incidentally, one may realize that the source plane and the lens plane defined 
based on the positions of the source and the lens in radial direction (or based 
on the distances of the source and the lens from the observer) have no relevance
with the variables in the lens equation.  The lens equation can only address the
relation between transverse position variables.   Then, one wonders why these 
objects play such a persistent role in the volitational papers and caustics. 

Let's pause for a moment and ponder angular variables.  What is it that we 
perceive and measure as the angular position of a celestial object?   
Speckle imaging may offer the best food for thought.   
The space time dependent refraction index makes a photon beam from 
a celestial object wiggle along through the atmosphere, 
and the object appears to hop around as registered at the focal plane.   
We may refer to these snap shots as a time series of ``apparent 
angular positions" of the object.  If we remove the atmospheric turbulence 
or the atmosphere altogether (again in our thought experiment) -- assuming 
that that leaves only the vacuum for the photon beam to propagate through,   
we will find a steady image on our CCD.  We may refer to it as the "true
angular position" of the object.   What is common for both ``apparent" and 
``true" angular positions is that an angular position is determined by the
direciton of the propagation vector of the photon beam at the observer.
The space time dependent refraction index of the atmosphere is an 
electromagnetic property of a matter in hydrodynamic motion.
On the other hand,  there is nothing wrong with understanding gravitational
lensing effect of the space time curvature in terms of effective refraction 
index (continuous function in space and time) assuming that we calculate
the index truthfully to the underlying physics.  That is, according to general
relativity, not in terms of Ferma theorem \cite{GL} unsubstantiated for the 
gravitational effect on optical paths (no historians have found evdience of
a margin the lack thereof prevented Ferma from elaborating Riemannian
geometry and propagation of massless spin one particles).    

Let's consider a quasar lens.
There are four objects that are believed to be all from the
same emission source QSO 2237+0305, and they are called Huchra's lens, Einstein
cross, or QSO 2237+0305.   The emission source QSO 2237+0305 is 
at a cosmological distance $z = 1.695$, and it will be a long time before
the lensing galaxy  moves away from the line of sight of the QSO even 
though the (Sb) galaxy is relatively close to us at $z = 0.0394$.   So, the four
objects are the thing that will be recognized as QSO 2237+0305 for generations to
come, but we always can remove the lensing galaxy in our thought experiment.
Then, an observer will see one object, say, at an angular position $\vec\alpha_s$,
which one may refer to as the ``true angular position" of the quasar.   
In a consistent manner, one might have liked to refer to the angular positions 
of the four objects as the ``apparent angular positions" of the quasar,  
but the multiplicity of the objects renders such practice seem unfit.   
The four objects can represnt the different facets of the lensed quasar,    
and they are usually referred to as the ``images" of  the quasar.  
So,  the angular positions of the four 
objects may be referred to as the ``angular positions of the images" of 
the quasar or simply the ``image positions".   This practice is in perfect 
harmony with our everyday experience.
A client in a barber chair next to a corner with mirrored walls can see  
multiple ``images" of oneself while being oblivious to the true object, oneself.
In fact, the images show the different sides of the client. 
So, we are all content to call the four objects the images of the quasar.
Now, we repeat our favorite thought experiment and reduce the mass of the
lensing galaxy to zero.  An observer will see one image of the quasar.  
What should we call this image?   Preimage?  Unlensed image? Image-sub-zero
(Image$_0$: the image one sees when the lensing mass vanishes)?  
Instead, this particular image is usually referred to as the ``source".
Then, the ``true angular
position" of the quasar $\vec\alpha_s$ may be referred to as the ``angular source 
position" or simply the ``source position", and equation (\ref{eqAlphaZeromass}) 
may read in English as follows:   the image position of an object is the
same as the source position of the object when there is no lensing mass. 
So, $\vec\alpha$ we introduced as the equivalent of $\zetavec/D_1$ is 
the variable for the image positions.  When we put back in the lensing galaxy,
the image positions may differ from the source position, and their relation
is nothing but the lens equation (\ref{eqLeqAlpha}).   One of the main games  
in quasar lensing (or any large scale lensing)  is to reconstruct $\vec\alpha_s$ 
and $\Thetavec$ from observational information of the images $\vec\alpha$.

In quasar lensing, the distance (actually the redshift) to the emission source
is one of the better determined quantities.   As a consequence, one may favor 
\cite{quasar_microlensing}  
to write the lens equation in terms of linear variables by multiplying the 
equation (\ref{eqLeqAlpha}) by the distance to the source $D_s$ (assumed to
be determined from the measured redshift and the cosmology to be determined). 
\opeqn
(D_s \vec\alpha_s) = (D_s \vec\alpha) - {D\over D_1} (D_s \Thetavec)
\label{eqLeqDs}
\cleqn
In the case of QSO 2237+0305, the variables ($D_s \vec\alpha_s$) and 
($D_s \vec\alpha$) may be considered to be defined on the plane at $z = 1.695$ 
from the observer.   (The equivalence between angular variables and
linear variables holds valid in a small scattering angle approximation,
$\Theta << 1$,  which we assume to be the case in this article. The angular
separations of the four objects in QSO 2237+0305 is about 
$1^{\prime\prime} << 1$~.)
One may refer to the plane at $z = 1.695$ as the source plane, 
as in G1421 (and references therein), and consider ($D_s \vec\alpha_s$) 
and ($D_s \vec\alpha$) as the linear variables projected into the source plane
\cite{quasar_microlensing}.    
Then, the variable $(D_s \vec\alpha_s)$ denotes the source position 
in the source plane, and the variable $(D_s \vec\alpha)$ denotes an image 
position in the source plane.   If one feels a bit of cluttered tautology here,    
one may realize that the culprit is the clinging desire to recognize the quasar
in full three-dimensional coordiantes of the space.  The source position has 
been assigned three coordiantes: $(D_s \vec\alpha_s, D_s)$, and so has 
been the position of the image: $(D_s \vec\alpha, D_s)$.   We note that the
images lie on the source plane  here.  We mentioned before that some practioners 
insist on putting images on the lens plane (or even lens planes).    

The source plane here is tied to the
value of the third (or radial) coordinates $D_s$ in $(D_s \vec\alpha_s, D_s)$
and $(D_s \vec\alpha, D_s)$.  On the other hand,  $\{D_s \vec\alpha_s\}$ spans 
a two-dimensional plane, and one may prefer to refer to the plane as the 
source plane because the plane is parameterized by the source position variable.   
The two source planes coincide and there doesn't seem to be any
conflict between the two definitions.  That is, until one realizes that
$\{D_s \vec\alpha \}$ also spans a two-dimensional plane, and one may prefer
to refer to the plane as the image plane because the plane is parameterized
by the image position variable.   So, we find ourselves in the middle of a 
luxury of definitions that  seem to be tangled in redundancy: the source plane 
originally  defined by the distance of the emission source at $D_s$ and 
parameterized by the (2-d) source position variable may be preferred to be referred
to as the source plane, and the source plane originally defined by the distance 
of the emission source at $D_s$ and parameterized by the (2-d) image position 
variable may be preferred to be referred to as the image plane.   

Lens equation is a relation of two diemensional variables and can accommodate 
only two dimensional angular variables or corresponding two dimensional
linear variables.  Even when one carries around the third (radial) components,  
the degrees of freedom in the lens equation is only  two dimensional.  One
can choose a plane that may represent the angular space faithfully with an
understooding that any plane is as good as any other plane.  
\begin{eqnarray*}
  (\vec\alpha_s, ~1) = D_s^{-1}~(D_s \vec\alpha_s, ~D_s)
       = D_\xi^{-1} ~(D_\xi\vec\alpha_s, ~D_\xi)  \\
  (\vec\alpha, ~1) = D_s^{-1}~(D_s \vec\alpha, ~D_s)
       = D_\xi^{-1}~(D_\xi\vec\alpha, ~D_\xi)
\end{eqnarray*}
One may choose the plane at a  unit distance, at $D_s$, or at an arbitrary 
distance $D_\xi$ from the observer.   They are all equivalent. 
The scattering angle is a quantity defined on an optical path 
-- a one-dimensional object in space.  Lensing is defined on the space of  
optical paths, which one as an observer recognizes only at one end of the paths.
Once a plane is chosen, that is where all the lensing variables will be defined
and compared.  Thus, it is best to consider the chosen plane as the 
``abstract plane" which may be parameterized by the source position variable 
or the image position variable.    We may prefer to refer to the ``abstract plane"
as the ``abstract lens plane"  or simply the ``lens plane"  because that is 
where the lens equation is defined and studied.  One may refer to the ``abstract
lens plane" parameterized by the source position variable as the source plane 
and the ``abstract lens plane" parameterized by the image position variable as 
the image plane.  So, the lens equation is a mapping from the ``abstract lens 
plane" to itself, or from the image plane to the source plane. 
\opeqn
(D_\xi \vec\alpha_s) = (D_\xi \vec\alpha) - {D\over D_1} (D_\xi \Thetavec)
\label{eqLeqDxi}
\cleqn

When $D_\xi = D_1$, the ``abstract lens plane" coincides with the lens plane
defined by the radial position of the lensing object as in  G1421 and ZG1199 
(and references therein).   As we will see in the following section, the
distance from the lensing mass of the apastron of an optical path around 
a lensing mass is the same as the Einstein ring radius on the plane at $D_1$ 
(in the approximation of linear gravity which is valid for all the observed 
and identified gravitational lenses).   Einstein ring refers to a ring image
(as well as the critical curve) in a point mass lens,  and there seems to be 
a (wrongful) religious belief among some practitioners that the plane at $D_1$
is endowed with a privileged position as the image plane.  That is not so, 
contrary to what one may find in G1421 and ZG1199.   It is indeed baffling to 
hear as recently as July 1999 a claim \cite{petters} that images can be in two 
different lens planes.  An observer does not see the photons in the images until 
they arrive at the observer as we have discussed repeatedly.  There is only one 
plane one may define:  the ``abstract lens plane", the representation of the 
observer's sky -- the motherboard of both the image plane and source plane.     
One can put the ``abstract lens plane" anywhere one finds it useful  as far 
as the smallness of the  scattering gurantees linearity between the plane and
the sky.    

The critical curve and the caustic curve pertain to the differential
beahvior of the lens equation.  The lens equation is an explicit mapping
from an image position to its source position, and there are multiple
solutions for a given source position.  The multiplicity can change from
domain to domain of the source plane, and all these interesting behavior 
can be studied starting from differentiating the lens equation. 
\opeqn
  d \vec\alpha_s = d \vec\alpha - {D\over D_1}
     {d \Thetavec \over d \vec\alpha}~ d\vec\alpha 
\label{eqDiff}
\cleqn
When one of the (never both in microlensing we are interested in)  
eigenvalues of this linear transformation, 
$d\vec\alpha \rightarrow d\vec\alpha_s$, vanishes, the lens equation is
said to be stationary along the eigendirection.   The set of the points
where the eigenvalue vanishes is called the critical curve.  This is a 
benign or natural generalization of what is familiar from a real function, 
say, $y = f(x): x , y \in {\cal Re}$. Critical points are where $df/dx =0$
(or f(x) is not differentiable).  Sometimes, we may hear "critical line"
in relation to lensing.  First of all, the set of critical points almost
always form closed smooth curves in lensing.  (One can assign lensing objects
at infinity and force the critical curve to have an open curve, whose physical
relevance I am not certain of.)   Also, ``critical line" may be
best left as the terminology for the loci of the zeros of the Riemann Zeta 
funcion.  (The zeros are believed to be on the ``line" whose real value is 
1/2, and this Riemann conjecture remains to be a conjecture despite the
telegram sent by Hilbert claiming otherwise.)   
Thus, we prefer to call the set of critical points of the lens
equation the critical curve.  The critical curve may be a disjoint sum of 
closed curves.   One may wonder what happens to the critical curve under
the mapping of the lens equation.  The resulting curve is called the caustic
curve.  The caustic curve has the same connectedness as the critical curve
because the lens equation is continuous (actually smooth) 
in the neighborhood of the critical curve.  
Continuity is preserved under a continuous mapping.   
On the other hand, I have no idea idea why the caustic
curve is named as it is,  even though they  look punky all right with 
spiky cusps.  In CRC Consise Encyclopedia of Mathematics,  caustics are 
defined as involutes, and involution vaguely reminds me of the way light rays
pile up on  the glittering surface of a swimming pool on a bright day.  
Caustic curves I encounter in microlensing are all ``some-form-of-oids" 
similar to cycloids: smooth closed curves punctuated by cusps.   
Cusps occur because the lens equation has stationary points along the critical 
curve.  A household name example of cusps may be the highest points in the 
swinging of a  pendulum.  The trajectory of the pendulum in space
changes the direction of the tangents (or the velocity) to the curve at
the stantionary points where the kinetic energy vanishes.  In The Random House 
College Dictionary, caustic is defined as to be severely critical, sarcastic,
or capable of burning living tissue.   We find the caustic curves relatively 
benign or even slightly enjoyable (we can generate relatively interesting looking
algebraic curves from physical necessity!), but the mythology around the caustic 
curve seems to have been, well, caustic.   One wonders whether the smoke in the 
backroom may be dully tributed as the shroud of acquired memories of Bruno burning 
at stake wondering of the neurochemistry of the minds of the inquisitors 
and leaving behind his philosophical conjecture on ubiquitous planets to be 
scientifically tested some four hundred years later.
 
So, where are the critical curve and the caustic curve?  They are objects 
defined through the lens equation and all lie in the same space:  angular space.  
Or, an equivalent linear space.     
We choose a lens plane, that is, an ``abstract lens plane", and mark
lens positions, source positions, image positions, Einstein ring, critical
curve and caustic curve.  And, anything else we may feel useful.  

Then, how does a caustic curve fly tachionically or at a speed faster than the 
speed of light?   It doesn't.  A caustic curve is not an object physically 
occupying a space at $D_s$ from the observer.  It does not swirl on the plane 
at $D_s$ from the observer in unison with a pair of binary masses in a relativistic 
orbital motion at a distance of $D_1$ from the observer.   
One may suggest:  apparent superluminal motion of blobs in a microquasar is 
a projection effect, and exactly the same phenomenon happens to the caustic 
curve once it is projected to a linear space such as the $D_s$-plane. 
Does it not?  Of course, not.  The analogy is flawed.   In the case of the 
blobs from the microquasar, they are the objects out there moving at certain 
linear velocities, and we can only measure the motions in terms of angular 
shifts in time.  The shifts of the angular positions of the blobs from the 
angular position of the microquasar which is the emission source of the blobs 
of particles can correspond to a speed larger than the speed of light when 
multiplied by the distance to the microquasar.  That is referred to as an 
(apparent) superluminal motion, and it offers an information on the direction of 
the beam of the blobs.  In contrast, the significance of defining superluminal 
caustic is as substantial as defining superluminal eyes 
upon having made a sweeping glance at the Milky Way from Ayers Rock.   
A caustic is nothing more than a peeping hole in this regard.   As the caustic
curve moves, the target stars that can be sampled through the window defined by 
the caustic curve change.  Furthermore, what we see are images not the caustic 
curve.  The caustic curve as an aperture does not deliver images directly.  
Images are delivered only after the tranformation dictated by the lens equation 
is carried out.   We will see in a following section that there is no 
``superluminality" to interest us even when we indulge in the phantom world
of tachyonic caustics.   Only the effect of fast glance: caustic crossing
signal buried under finite size source effect and ``long" exposure.

\begin{itemize}
\item
  \underline{\Large Lens Equations are Relations of Optical Paths}
\end{itemize}

Schwarschld metric is an exact solution to the Einstein field equations
with a point mass.   If the point mass is $M$,  the Schwarzschild metric is 
given by  \cite{sweinberg}
\opeqn
 ds^2 = -\left(1-{2GM\over r}\right) dt^2
        + {dr^2\over 1 - 2GM/r} + r^2 (d\theta^2 + \sin\theta^2 d\phi^2) \ ,  
\label{eqMetric}
\cleqn
where the Schwarzschild radius $r_s = 2GM$ is $2.95$km for a solar mass
object.    In microlensing,
photons' passage is about $10^8$ times the Schwarzschid radius.
The optical path is found by solving the free fall equation with 
the null condition $ds^2 = 0$, and
the orbit $\theta(r)$ is given as an elliptic integral that requires numerical
estimation.  When the closest
approach $r_{\circ}$ of the photon beam to the mass $M$ is $10^8$ times $r_s$
or so, however, the weak gravity allows truncation of the integral at the linear 
order (in the Newtonian potential $GM/r$, which is called Robertson expansion), 
and the scattering angle of the orbit takes a simple form. 
\opeqn
   \Theta =  {4 G M \over r_{\circ}}  = {2 r_s \over r_{\circ}} 
\label{eqTheta}
\cleqn
In Newtonian gravity, the scattering angle is given by twice the value of
the Newtonian potential at the closest approach, $2GM/r_{\circ}$,
and differs from Einstein gravity factor 2.  This factor 2 difference 
was crucial in establishing Einstein theory as the theory of gravity.    

In Newtonian gravity, an unbound orbit forms a hyperbolic curve 
(on the plane defined by an azimuthal angle $\phi =$ constant).
If we consider the family of hyperbolic curves connecting two asymptotic
points that represent an emission source and an observer, the scattering
angle the hyperbolic curve represents grows with the distance from
the lensing mass located somewhere between the emission source and      
the observer.  In GR, the photon trajecotries (in the Schwarzschild coordinates) 
are   not exactly hyperbolic, but the family of photon trajectories
 share the same behavior:
the scattering angle grows with the distance to the lens position.  
 On the other hand, equation (\ref{eqTheta})  tells us that
the scattering angle $\Theta$ is inversely proportional to $r_{\circ}$.   
Therefore, there are only two possible null geodesics from a given emission 
source to a given obserser for a given azimuthal angle.   

\begin{figure}
\plotone{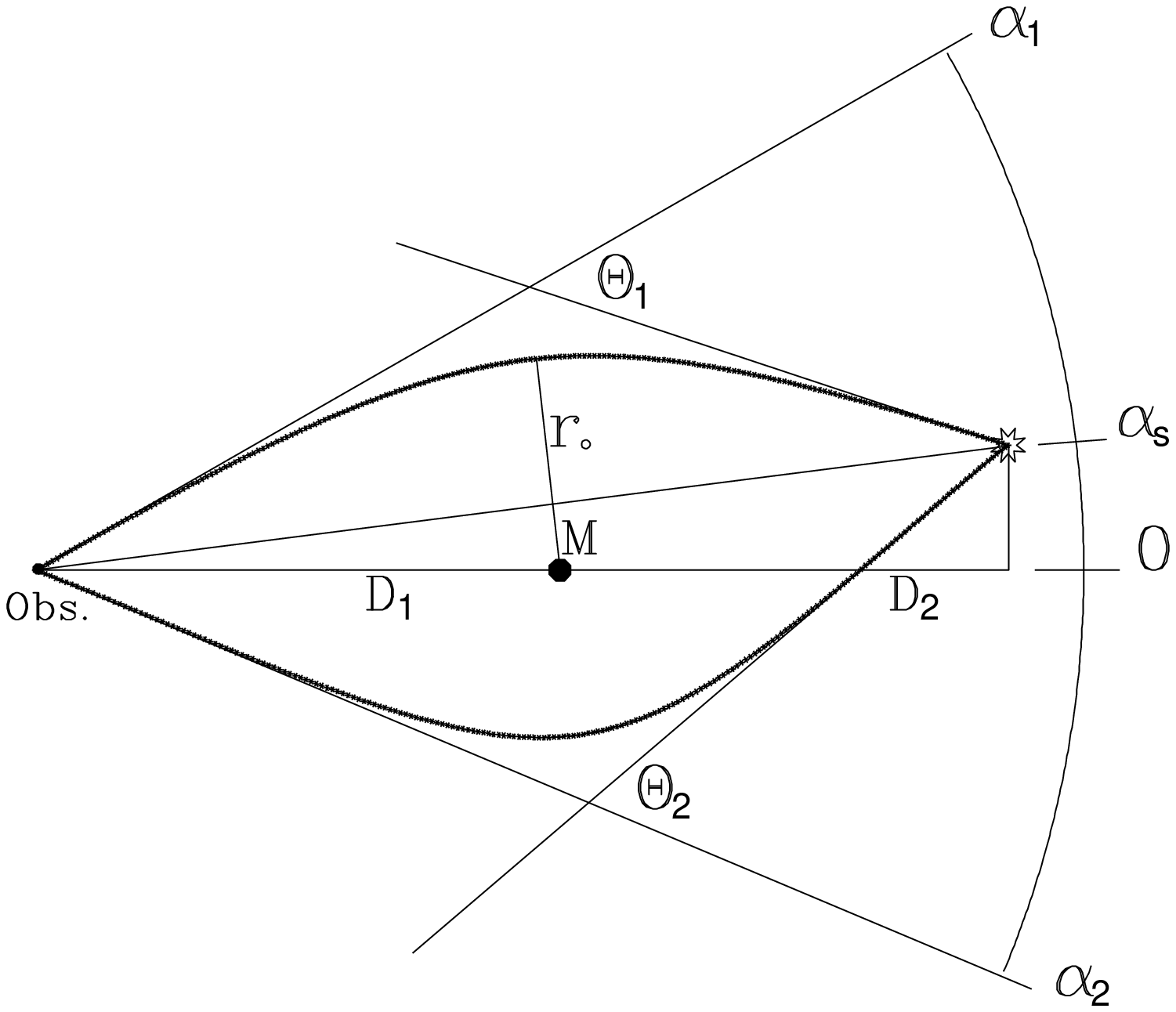}
\figcaption{\label{fig-scatplane}
}
\end{figure}

Figure \ref{fig-scatplane} shows the scattering plane ($\phi =$ constant) 
and two null geodesics (or optical paths).  A photon emitted along the tangent
to an optical path at the emission source arrives at the observer with
the propagation vector tangent to the optical path at the observer. 
Thus, the observer sees two stars, one at $(\alpha_1, \phi)$ and the other
at $(\alpha_2, \phi)$ (in this unorthodox angular position coordinate system).    
If we remove the lensing mass $M$ (or wait for the lensing mass to move away), 
the observer will see one star at $(\alpha_s, \phi)$.   In lensing jargon,
$(\alpha_s, \phi)$ is referred to as the source position, and $(\alpha_1, \phi)$ 
and $(\alpha_2, \phi)$ are referred to as the image positions.   The relation 
between the source position and the image positions is called the lens 
equation and is obtained easily from the diagram in figure \ref{fig-scatplane}.
If  $\alpha$ is the variable for the image positions, and $D_1$ and $D_2$
are the distances from the lens to the observer and to the source along the line
of sight, the lens equation is  given by 
\opeqn 
  \alpha - \alpha_s = {D_2\over D_1 + D_2}~{4GM\over D_1 \alpha}  
         \equiv  {\alpha_E^2 \over \alpha} \ .  
\label{eqAleq}
\cleqn
This is a quadratic equation and has two solutions for $\alpha$ for each $\alpha_s$.
When $\alpha_s =0$, the two solutions are reflection symmetric:  
$\alpha = \pm\alpha_E$.  In fact, when $\alpha_s =0$, the scattering plane is 
not uniquely determined due to the azimuthal symmetry, and the images form 
along a ring  of radius $\alpha_E$.   This ring is the famous Einstein ring, 
and $\alpha_E$ is referred to as the angular Einstein ring radius.   In the small 
scattering angle approximation we are using,  the closest approach $r_\circ$
of these photon paths to the lensing mass $M$ is the same  as the Einstein
ring radius $R_E \equiv D_1 \alpha_E$.  
$D_1$ and $D_2$, 
\opeqn
   R_E \equiv \sqrt{4GMD}  \ ,   
\cleqn
where $D$ is the reduced distance of $D_1$ and $D_2$.
The lens equation (\ref{eqAleq}) can be written in terms of linear variables.
Let $b \equiv D_1 \alpha$ and  $s \equiv D_1 \alpha_s$.  Then,  the variables
are defined  on the plane that passes through the lensing mass.
\opeqn
  b - s = R_E^2~ {1\over b} 
\cleqn
In order to incorporate the variable ($\phi$) for the orientation of the 
scattering angle,  we should write it as a (2-d) vector equation.
\opeqn
 \vec b - \vec s = R_E^2~ {\vec b\over {\vec b}^2}
\label{eqSingb}
\cleqn
So far, the lensing mass has been at the origin of the coordinate system.
If it is at $\vec x$,  the lens equation becomes
\opeqn
 \vec b - \vec s = R_E^2 {\vec b - \vec x \over (\vec b - \vec x)^2} \ . 
\cleqn
This can be extended to multiple particle lens systems.  For a binary lens,
\opeqn
 \vec b - \vec s 
   = R_E^2~ \left({\epsilon_1(\vec b-\vec x_1) \over (\vec b-\vec x_1)^2} 
      + {\epsilon_2(\vec b-\vec x_2) \over (\vec b-\vec x_2)^2}\right)  \ ,
\label{eqBib}
\cleqn
where $R_E$ is the Einstein ring radius of the total mass $M$,
$R_E^2 \equiv 4 G M D$,    and $\epsilon_1$ and $\epsilon_2$ are the fractional 
masses located at $\vec x_1$ and $\vec x_2$ respectively.  

The two-dimensional vectors are most ideally handled as complex variables.
Most of all,  that is the only way to solve the binary equation.
So, let's complexify the variables: $\vec b, \vec s, \vec x_1 \vec x_2 
\rightarrow z, \omega,  x_1, x_2$.   Then equations (\ref{eqSingb})  and
(\ref{eqBib}) are rewritten as follows.
\opeqn
   \omega = z - {R_E^2\over \bar z} 
\label{eqSing}
\cleqn
\opeqn
   \omega = z - R_E^2 \left({\epsilon_1 \over \bar z - x_1}  
                     + {\epsilon_2 \over \bar z - x_2} \right)
\label{eqBi}
\cleqn
We can choose the coordinate system so that the lens position variables
$x_1$ and $x_2$ are real.   Once we introduce a complex variable on the plane
on which the lensing variables are defined  (so commonly referred to as the
lens plane), the lens plane as a two-dimensional linear space is parameterized
by the complex variable and its complex conjugate.   What is convenient about
complex variables is that we only need to write half the equation.   For example,
equation (\ref{eqSing}) implies that the following is also true.
\opeqn 
   \bar\omega = \bar z - {R_E^2\over z}
\cleqn
Incidently, we have defined two sets of variables on the lens plane:  One
for the image position variable, $(z, \bar z)$, and  the other for the source 
variable position variable, $(\omega, \bar\omega)$.  One may wonder if it is 
necessary to consider a projected plane to define complex variables.  That 
is not so.  We could have defined a complex plane for the angular space 
parameterized by $\vec \alpha (\leftarrow \alpha)$ or $\vec \alpha_s
(\leftarrow \alpha_s)$.    It is just that we have identified  the angular 
space and a projected plane,  which is valid because $\Theta << 1$.  As a 
matter of fact, we could have chosen any projected plane as our lens plane
where we define lensing variables.   What is invariable is the fact that 
the observer sees images and recognizes the tangent of the optical paths
at the observer as the angular positions of the images in the sky. 
In fact, we do not have to adhere to the linear scale of the projected plane,
and it is customary to normalize the equation  (or scale the lens plane) 
so that $R_E = 1$.   For example, the binary lens equation 
can be rewritten in dimensionless variables as follows.      
\opeqn
  \omega = z -  {\epsilon_1 \over \bar z - x_1}
                     - {\epsilon_2 \over \bar z - x_2}    
\cleqn

\begin{itemize}
\item
  \underline{\Large  Off-Axis Trioids of a Binary Lens}
\end{itemize}

Einstein ring is well known, but it is still an interesting object to think
about if we think about it.   If the radiation emission from the source is a
uni-directional coherent beam as in a laser,   the observer will be able to 
see one image of the source at most let alone a ring image.   However, many
heavenly bodies are largely isotropic radiation emitters.  So, when the  
lensing trio are aligned, the observer sees infinitely
many rays from the source, $\{\vec\alpha~|~ |\vec\alpha| = \alpha_E\}$, instead 
of one, $\{\vec\alpha~|~ \vec\alpha = \vec\alpha_s \}$.   
If we look at the lens equation 
(\ref{eqSing}), it is an explicit mapping from an image position to its
source position.  So, when the lensing trio is aligned, a continuum of image 
positions is mapped to one source position under the lens equation.  In other
words,  the Einstein ring is the set of statinary points of the lens equation.
The curve of stationary points of a mapping seems to be said to be critical
(I have an impression that anything in mathematics that may be remotely interesting 
is said to be critical), 
hence Einstein ring is a critical curve of a point mass lens.   The stationarity
is due to the azimuthal (or axial) symmetry of the lensing system, hence the
tangent (azimuthal vector) to the Einstein ring vanishes under the lens equation 
but not the normal.  The linear differential behavior of a mapping is conveniently
described by the Jacobian matrix ($d\vec\alpha_s/d\vec\alpha$) written out in
a 2 by 2 array, and the criticality shows up as a vanishing
eigenvalue of the Jacobian matrix.   We differentiate equation (\ref{eqSing}).
\opeqn
     {d\omega \choose d\bar\omega} =
        \pmatrix {\del_z\omega & \del_{\bar z}\omega \cr
                       \del_z\bar\omega & \del_{\bar z}\bar\omega  \cr}
        {dz \choose d\bar z}
     \equiv \ {\cal J} \  {dz \choose d\bar z}
\label{eqDerivative}
\cleqn
where the Jacobian matrix ${\cal J}$ is given as follows.
\opeqn
 {\cal J } = \pmatrix {1    &    \bar\kappa   \cr
                       \kappa    &    1   \cr} \qquad ; \qquad
             \kappa \equiv {R_E^2\over z^2}
\label{eqJac}
\cleqn
The eigenvalues  are
\opeqn
 \lambda_{\pm} = 1 \pm |\kappa| \
\label{eqEval}
\cleqn
(The eigenvalues are real, which must be expected from that
the Jacobian matrix is hermitean : \ ${\cal J}^{\dagger} = {\cal J}$).
When $|z| = R_E$, $|\kappa| = 1$, and $\lambda_-$ vanishes.  So does the 
Jacobian determinant, of course, which is the product of the eigenvalues.

\begin{figure}
\plotone{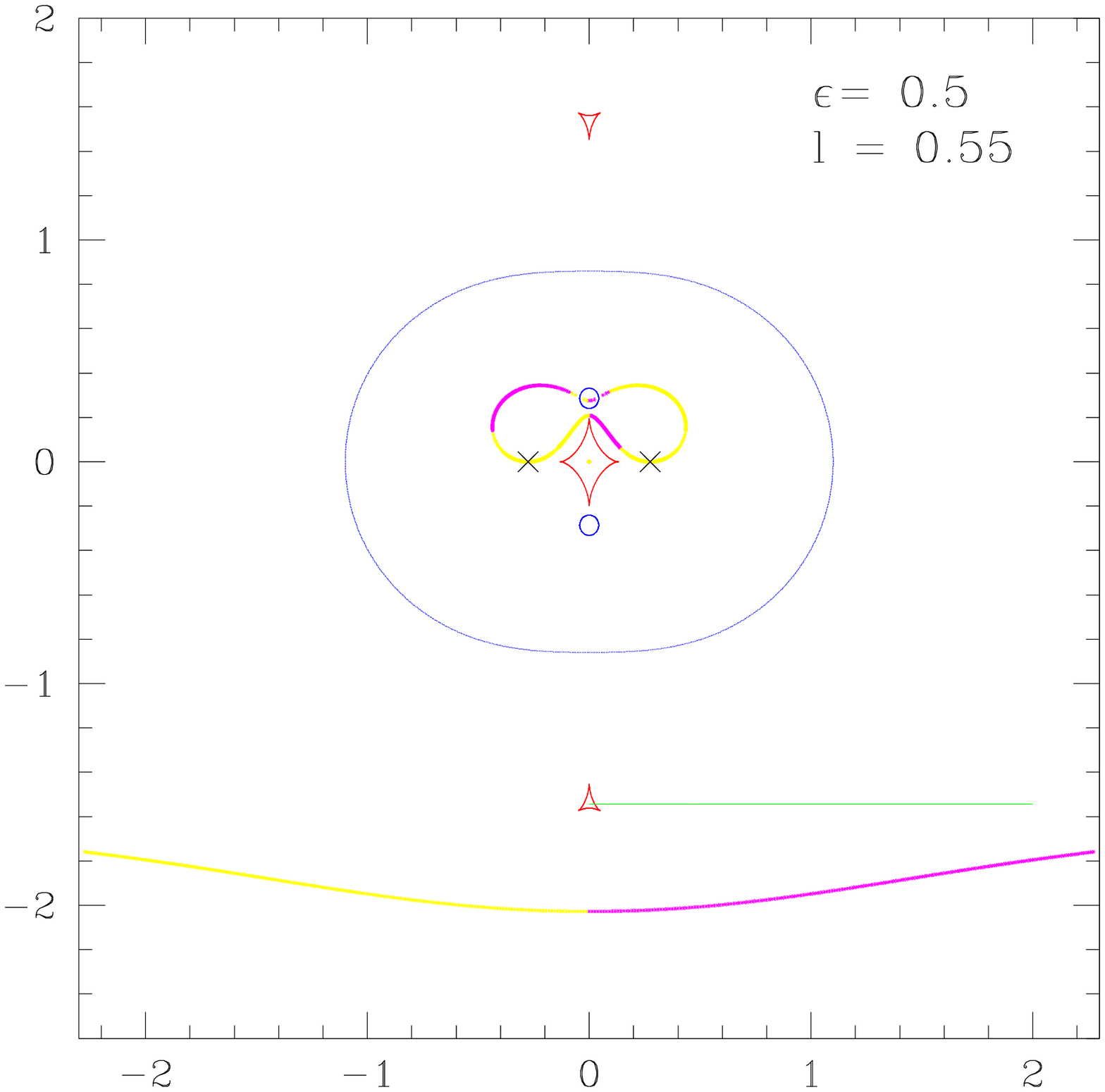}
\figcaption{\label{fig-caustic}
}
\end{figure}

In the case of a binary lens, the differential equations are exactly the
same except for that $\kappa$ is given by 
\opeqn
 \kappa \equiv {\epsilon_1\over (z-x_1)^2} +
               {\epsilon_2\over (z-x_2)^2} \ . 
\label{eqKappa}
\cleqn
We have chosen $R_E = 1$ as in the (normalized) binary equation (\ref{eqBi}).  
On the critical curve, where $\lambda_- = 0$ and $\lambda_+ = 2$, $\kappa$ 
is a pure phase because $|\kappa| = 1$.   So, the critical curve is the set
of the solutions to the analytic equation (\ref{eqKappa}) with 
\opeqn
   \kappa = e^{2i\varphi} \ : \qquad \varphi \in [0, \pi) \ . 
\cleqn   
When the separation $\ell \equiv |x_1 - x_2|$ between the two lens elements
is smaller than $\ell_-$,  the critical curve is made of three loops and so
is the caustic curve.  
\opeqn
  \ell_- \ = \ \left({\root 3 \of {\epsilon_1}}
                   +{\root 3 \of {\epsilon_2}}\right)^{-{3\over 4}} \ \ ;
       \qquad {1 \over \sqrt{2}} \ \le \ \ell_- \ < \ 1
\cleqn 
So, the caustic curve of a binary lens with $\ell \lsim 0.7$ is made of three
disjointed loops irrelevantly of the fractional mass parameter $\epsilon_s$.    
Figure  \ref{fig-caustic} shows an example of a (symmetric or equal mass) binary 
lens with $\ell = 0.55$.   The critical curve is in blue, the caustic curve is 
in red, and the two crosses in black are the positions of the lenses.  The line
that connects the two lens elements of a binary lens is referred to as the lens
axis.  The caustic loop with four cusps (quadroid) always crossed the lens 
axis, and the two triangular caustic loops (trioid) alway are off the lens axis.
The small critical loop encloses the limit point which is at 
$z_{\ast\pm} = \pm i \ell/2$.  The corresponding points (correspondence by the 
mapping of the lens equation) fall inside the trioids.
\opeqn
  \omega_{\ast\pm} = \pm i \left({\ell\over 2} - {1\over \ell}\right)
\label{eqTriPosition}
\cleqn 
Figure \ref{fig-caustic} shows  a source trajectory with
one end at $\omega_{\ast +}$  in greeen and the corresponding 
image trajectories  in magenta.   We have chosen this half-way trajectory 
so that the accidental symmetry due to the equal mass would not mire the visual
clarity of the behavior of the image trajectories.   The union of the yellow 
curves and the magenta curves represent the total images of the line source 
trajectory with $\omega = - 1.75 i$.   Readers are encouraged to be impressed
by the similarity between the source trajectory and the image trajectory at 
the bottom of the plot and the relatively small area the image trajecotries
inside the large critical loop occupy.   The parity of the images is positive
outside the large critical loop and inside the small critical loops.  Inside
the large critical loop and outside the small critical loops, the images have
negative parity.    There are usually three images in a binary lens,  and the
corresponding image trajectories are the one  at the bottom of the plot converging
to $\infty$ (positive) and the two outside the small critical loop converging
to the lens positions.   Since $\omega = - 1.75 i$ is inside a caustic loop, 
we expect two more images while the source trajectory remains inside the caustic
loop.   They are the small segments that connected at a critical point on the
small critical loop in the upper half plane.         
 
In the case of a symmetric lens, the two limit points and the lens positions
form a square.   So, the distance between the two small critical curves is 
about the same as the separation between the lensing masses, and the caustic
crossing images are at a distance of about half the separation from the center 
of the mass.    As $\ell$ becomes small,  the distance of the limit points 
decrease linearly with $\ell$ and do all the images but the image near the source.    
This means that all the images but the image near the source become only nominal
images, and that is reflected on the off-axis caustics moving away from the 
lens axis inversely proportional to the separation $\ell$  (see 
equation (\ref{eqTriPosition})).  Also, the sizes of the critical loops and 
caustic loops shrink as $\ell$ shrinks.    What it means is that the lensing
elements are so close to each other that the binary lens behave  more or
less as a single lens.   If we assume that $\ell = 0.1$ as in ZG1199,  the 
caustic in the lower half-plane will be at $\omega = - 9.95 i$.   The microlensing 
amplitude of a single lens of a source at a distance of 9.95 (in Einstein ring 
radius unit) from the lens  is  1.000196.  So,  the effect of lensing
on the image near the source is  a brightening by $\sim 0.02 \%$, which is 
practically equivalent to no lensing at all.   Also, as the trioid caustic
shrinks practically to a point at $\omega = - 9.95 i$,  the finite size
source effect washes out the singular brightening effect of the critical 
curve (the images of the most of the part of the star falls away from the
critical curve, and its average falls below any reasonable detection level).    
The side of the trioids measure about 0.00045 in units of Einstein ring radius.
Usually, the Einstein ring radius is $\lsim 1$ mas,  then the size of the troids
will be $\lsim 0.45 \mu$as.  The solar radius at 8 kpc from us will  be about
0.565 $\mu$as. 

Now, let the binary rotate.  Let's put the (abstract) lens plane at $D_1$.
The linear speed of the troids will be hundred times larger than that of the 
binary masses.  If we assume that the lens is half way to the source  that
is 8 kpc away from us, then the Einstein ring radius of a solar mass lens 
is 4 au.  If the binary is face on, and $\ell = 0.4$ au, then the orbital
velocity is $50$ km/sec, and trioids move at an apparent speed of 
$5,000$ km/sec $= 0.0167$ (in units of the speed of light) which is hardly 
a relativistic speed.     If the lensed star has the solar radius, then it 
is 1.16 seconds on the (abstract) lens plane.  So, it takes about 140 seconds 
for the trioid to sweep across the solar diameter.  If the  exposure time is 
order of a few minutes (with a moderate size telescope), the signal of the 
caustic crossing will be contained in one frame.  If one arranges the apparent
speed of the trioid to be bigger, the  effect will be to shorten the duration
of the signal.    There is no physical bearing  this ``fast glance effect"
of a far-away caustic shares with the ``superluminal effect" of a particle beam 
moving at an angle with respect to the line of sight.

\begin{itemize}
\item
  \underline{\Large  Name That ``-OID"}
\end{itemize}
 
In a gravitational binary lens, one encounters three types of 
``some-form-of-OID's":  ``trioid" with three cusps, ``quadroid" with
four cusps, and ``hexoid" with six cusps borrowing the names from our own 
paper on line caustic crossing and limb darkening \cite{limbpaper}.  
The ``-OID's" in binary lenses
are all simple loops with winding number one unlike in higher 
multiple point lenses.  In an effort to avoid cooking up redundant 
nomenclatures for ``some-form-of-OID's", we looked up  
``CRC Concise Encyclopedia of Mathematics" edited by Chapman and Hall (CRC from
hereon) with ``tricuspid" suggested by an ``authority" for ``trioid" in mind.

``TricuspOID" seems to be a mathematical term even though ``tricuspid" is not.
So is  ``deltoid", which seems to originate from the shape of the Greek letter
$\Delta$.  It is also the anatomic term for a large muscle covering the
shoulder joint.   We get an impreesion that anatomy and geometry must have been
developed hand in hand.   A ``nephroid" is an ``-oid" with two cusps one 
can generate using a so-called supercritical lens.  It looks somewhat ``like a
kidney" (again from Greek), and so its name, ``nephroid".  The cusps of a 
``nephroid" are spiky inward, hence a ``nephroid" is an epicycloid. It doesn't
seem to be a taboo  to refer to a   ``nephroid" as a ``2-cusped epicycloid".   
An epicycloid with one cusp resembles the heart, and so a ``cardioid".  The  
parametric equation seems to be easy to recognize due to the close relation 
to the polar equation for an ellipsis. 
It is $r= a(1+\cos\theta)$ for a ``cardioid" and $a= r(1+\cos\theta)$ for an
ellipsis (with eccentricity 1).  We have failed in finding anatomic names for
epicycloids with three cusps or more.  However,  we have found a stellar
nomenclature for a 4-cusped object.  An ``astroid" is a hypocycloid (spiky
outward) with four cusps.   It is also called a tetracuspID, cubocycloid,
or paracycle according to our reference CRC.   Cubocycloid must have derived from 
cuboid which refers to a rectangular parallelepiped and also one of the tarsal 
bones.   

In microlensing where the lenses are gravitationally bound multiple point masses, 
the metric is flat asymptotically, and  the image of a source far away from the
lens system is an unlensed image (or source itself) with $J = 1$.  The critical
curves are always closed curves, and the lens equation is always subcritical 
even when one includes dispersed medium of Galactic particle dark matter.
One consequence is that the caustic curves are smooth closed curves punctuated 
by (spiky outward) cusps, or simply, hypocycloids.   There is no 1-cusped 
hypocycloid or 2-cusped hypocycloid.  After consulting the 2000-page reference, 
we may still feel in jeopardy how to  extend the naming tradition of ``-OID's". 
Or, is it ``-ID's"?   The confusion between ``-oid" and ``-id" can arise from 
that a 3-cusped hypocycloid is called a ``tricuspOID", and a 4-cusped 
hypocycloid is called a ``tetracuspID" according to CRC.  Considering that 
a 4-cusped hypocycloid is most commonly  referred to as an ``astroid" not
``tetracuspid",  we conjecture that  ``tetracuspid" must have derived from
dental teminology ``cuspid" and lapsed attention to the particular 
characteristics of the points (cusps) in cycloids.   
A cuspid refers to a canine tooth which has a single projection point.
A bicuspid refers to a premolar tooth which has two projection points.
A tricuspid refers to a tooth that has three projection points and also
a tricuspid valve.   The suffix ``-cuspid" seems to mean ``pointed". 
Cusps in cycloid are pointed in a particular manner where the tangent flips
its sign.   This explains why ``tricuspid" is not a mathematical terminology
for a 3-cusped hypocycloid.   This explains why we referred to these objects
as ``some-form-of-oids" early on in this article.  There doesn't seem to be 
a usage of tetracuspid in dentistry or in anatomy.    

Now, we discuss why chose the pattern of ``number-oid" (``quadroid") instead
of ``number-vertex-oid" (as in ``tricuspoid") or ``anatomy-oid" (as in
``nephroid").   Following the ``tradition" of borrowing anatomical names  
is excluded because of the likelihood of arbitrary  number of cusps that may
define caustic curves.     So, the extension of imagination that may stem from 
``deltoid" and ``cubocycloid" (and also ``astroid" which is obviously an anatomy
of a star if we look at a bright star in an HST frame) meets the dead end.  
We read that Euler studied a deltoid
in 1745 in relation to an optical problem and also by Steiner in 1856, and a
deltoid is referred to as Steiner's hypocycloid in some literature.  Despite all 
our intention to honor them, our option becomes limited to the pattern of
``number-oid" or ``number-vertex-oid".  
Let's consider epicycle and epicycloid to differentiate the two.  
An epicycle depicts a circular motion of an object around a center that 
moves along a larger circle,  and  the curve is smooth everywhere.  
An epicylcoid depicts a cicrular motion of an object around the smaller circle  
that rolls on the larger circle, and the curve is smooth except at the cusps
(the winding number is determined by the ratio of the two circles).  It seems
to be clear that ``-oid" in ``-oid" objects we discussed so far represents 
a particular resemblance to a cycle: curved and smooth like a circle but 
punctuated with points where the tangent flips its sign.  In a binary lens, 
they are also simply closed curves.  Thus, it is very clear that the number of 
cusps and the number of the smooth segments (or ``sides") of a caustic loop 
are the same, and it is sufficient to assign a number to specify the particular
``-oid".   In a higher multiple point lens,  a caustic loop can have winding
number larger than 1.  However, the winding number is a finite integer, and 
the number of vertices (or cusps) and the number of sides (or smooth segments)
are the same.  The caustic curve of a gravitational binary lens consists of 
one hexoid, two quadroids, or, one quadroid and two trioids.

\begin{itemize}
\item
  \underline{\Large  Preferences and Censors}
\end{itemize}

We have reviewed that the intrinsic variables of the lens equation are
angular variables.  As in figure (\ref{fig-caustic}), we are free to choose
a plane, set the distance scale, and mark all the variables, parameters, and
objects we find fit from the lens equation.  When $\ell \rightarrow 0$, the
binary lens converges to a single lens, and the quadroid caustic around the 
center of mass contracts to a point caustic.  So, ``the single lens has a 
point caustic at the position of the lens."  We may draw the critical curve
and caustic curve on the same (abstract) lens plane as we did in figure
(\ref{fig-caustic}). The lens equation is a mapping from the chosen (abstract)
lens plane to itself, or a mapping from the image plane to the source plane
where the image plane refers to the (abstract) lens plane parameterized by the
image position variable and the source plane refers to the (abstract) lens plane
parameterized by the source position variable.  

Now, is it confusing to call $D_s$-plane (the plane defined by the radial position
of the radiation emission source) the source plane?  Having understood that we
only need to define one space (angular space) or a plane that corresponds to the
angular space, it doesn't seem to matter whatever the plane may be called. 
Once we know clearly what degrees of freedom we are manipulating through the
lens equation, it doesn't seem to be a  confusing practice to let the beloved
term ``source plane" be used in both ways: based on the radial coordinate or
based on the transverse coordinates.  What is invariant seems to be that physics
lies in relations not in nomenclatures.  It is good to have distinguishable
nomenclatures, but there is nothing holy (meaning leaving no room for scientific
reasoning and fluidity) about the ``source plane". We find it sufficient to 
exercise a bit of contextual understanding to let the ``source plane" enjoy the
both definitions and let lensing colleagues keep their inertial tradition of 
describing lensing.  Is it a law or a rule that the caustic curve lies on the
source plane?  We may choose to call the plane parameterized by the source position
variable $\omega$ the $\omega$-plane, perhaps from boredom or from a courtesy
to leave the term ``source plane" for the practitioners who are attatched to
the radial coordinate.  Then, the caustic curve will lie on the $\omega$-plane.
So, it may constitute a proper question to ask if it is a law  or a rule, and
no paper should be shredded over this unsubstantiated dogma on wording. 

Are newcomers to the field confused by the confusing usage of terminologies
as Gould claims?  Perhaps, not. We find it an unsubstantiated claim.   Considering 
the suggestion of superluminal caustics, we would think that the proper 
route to resolve any confustion is for students to take a bit more than a few 
hours to understand lensing from first principles.   We have found that the 
origin of the controversies lies in conceptual misunderstandings.

\begin{itemize}
\item
  \underline{\Large  Rapidly Rotating Black Holes and Optical Paths}
\end{itemize}

Maxwell's equations tell us how electromagnetic fields and matter interact.
Lienard-Wiechert potentials satisfy the Maxwell's equations with a moving
charge.  Special relativity is a property of the space time and so is the 
spin of a particle.  When the velocity of the system becomes comparable to 
the speed of light,  the space-like components of the fields become comparable
to the time-like components.  And, a novelty one may witness (as a student)
from playing with the Lienard-Wiechert potentials is to see the electromagnetic 
wave propagate and actually carry the energy out to infinity.  
If we only pick out the time-component
$A_0$, we will be at a loss with the discrepancies with the measurements of the
electric and magnetic fields.   Einstein field equations tell us how gravitational
fields (or metric) and matter interact.  When the velocity of the system becomes
comparable to the speed of light, we expect that one should examine not only the
time-time component ($g_{00}$) but also the other five components of the metric. 
We find it a baffling practice for ZG1199 to declare without substantiation that 
``retarding the Newtonian potential" (with additional factor 2 mentioned above) 
results in the ``gravitational analog" that governs the behavior of the null 
geodesics as seen by an observer.    We have no doubt that it will take a bit
more than a few hours to examine the metric even at the  post Newtonian level. 
However, we find it a worthy exercise to be carried out.

\acknowledgements
\section*{Acknowledgments}

It is our pleasure to express our gratitude to Clara Bennett for the copy of 
``CRC Concise Encyclopedia of Mathematics" given to the author during the  
last winter solstice.  It is a great gift to be snowed in with.

% references here

\def\ref@jnl#1{{\rm#1}}
\def\aj{\ref@jnl{AJ}}
\def\apj{\ref@jnl{ApJ}}
\def\apjl{\ref@jnl{ApJ}}
\def\apjs{\ref@jnl{ApJS}}
\def\aap{\ref@jnl{A\&A}}
\def\aapr{\ref@jnl{A\&A~Rev.}}
\def\aaps{\ref@jnl{A\&AS}}
\def\mnras{\ref@jnl{MNRAS}}
\def\prl{\ref@jnl{Phys.~Rev.~Lett.}}
\def\pasp{\ref@jnl{PASP}}
\def\nat{\ref@jnl{Nature}}
\def\iauc{\ref@jnl{IAU~Circ.}}
\def\aplett{\ref@jnl{Astrophys.~Lett.}}
\def\annrev{\ref@jnl{Ann.~Rev.~Astron.~and Astroph.}}

\clearpage

%\onecolumn

% \label{}
% \end{tabular}
% \end{table}

\end{document}